\documentclass[aps,prl,twocolumn,superscriptaddress,amsmath,amssymb]{revtex4}
\usepackage{graphicx,overpic} %
\usepackage{epstopdf} %
\usepackage{dcolumn} %
\usepackage{xcolor} %
\usepackage{hyperref} %
\usepackage[utf8]{inputenc} %
\usepackage[english]{babel} %
\usepackage[displaymath, mathlines]{lineno} %
\usepackage{blindtext} %
\usepackage{amsmath} %
\usepackage{amssymb}
\usepackage{amsfonts}
\usepackage{upgreek} %
\usepackage{booktabs}
\usepackage[normalem]{ulem}

\newcommand*\patchAmsMathEnvironmentForLineno[1]{%
\expandafter\let\csname old#1\expandafter\endcsname\csname #1\endcsname
\expandafter\let\csname oldend#1\expandafter\endcsname\csname
end#1\endcsname
 \renewenvironment{#1}%
   {\linenomath\csname old#1\endcsname}%
   {\csname oldend#1\endcsname\endlinenomath}%
}
\newcommand*\patchBothAmsMathEnvironmentsForLineno[1]{%
  \patchAmsMathEnvironmentForLineno{#1}%
  \patchAmsMathEnvironmentForLineno{#1*}%
}
\AtBeginDocument{%
\patchBothAmsMathEnvironmentsForLineno{equation}%
\patchBothAmsMathEnvironmentsForLineno{align}%
\patchBothAmsMathEnvironmentsForLineno{flalign}%
\patchBothAmsMathEnvironmentsForLineno{alignat}%
\patchBothAmsMathEnvironmentsForLineno{gather}%
\patchBothAmsMathEnvironmentsForLineno{multline}%
\patchBothAmsMathEnvironmentsForLineno{eqnarray}%
}

\graphicspath{{figures/}} %

\usepackage{relsize}
\def\babar{\mbox{\slshape B\kern-0.1em{\smaller A}\kern-0.1em
    B\kern-0.1em{\smaller A\kern-0.2em R}}\xspace}

\newcommand{\eg}{\mbox{\itshape e.g.}\xspace}
\newcommand{\ie}{\mbox{\itshape i.e.}\xspace}

\newcommand{\vs}{\mbox{\itshape vs.}\xspace}

\newcommand{\lumi}{\ensuremath{72\invfb}\xspace}

\newcommand{\DpToKpipi}{\ensuremath{\Dp\to\Km\pip\pip}\xspace}
\newcommand{\DzToKpi}{\ensuremath{\Dz\to\Km\pip}\xspace}

\newcommand{\mDKpi}{\ensuremath{m(\Km\pip)}\xspace}

\newcommand{\mDp}{\ensuremath{m(\Km\pip\pip)}\xspace}
\newcommand{\dm}{\ensuremath{\Delta m}\xspace}
\newcommand{\sigmat}{\ensuremath{\sigma_t}\xspace}

\newcommand{\tauDzKpi}{410.5}
\newcommand{\tauDzStatKpi}{1.1}
\newcommand{\tauDzSystKpi}{0.8}

\newcommand{\tauDzKpipipi}{408.8}
\newcommand{\tauDzStatKpipipi}{1.2}

\newcommand{\tauDz}{\tauDzKpi}
\newcommand{\tauDzStat}{\tauDzStatKpi}
\newcommand{\tauDzSyst}{\tauDzSystKpi}

\newcommand{\tauDp}{1030.4}
\newcommand{\tauDpStat}{4.7}
\newcommand{\tauDpSyst}{3.1} 
\RequirePackage{xspace}

\newcommand{\nospaceunit}[1]{\ensuremath{\text{#1}}}       

\def\belletwo{Belle~II\xspace}

\newcommand{\CP}{\ensuremath{C\!P}\xspace}

\def\epem       {\ensuremath{e^+e^-}\xspace}

\def\mup        {\ensuremath{\mu^+}\xspace}
\def\mun        {\ensuremath{\mu^-}\xspace} %

\def\c     {\ensuremath{c}\xspace}
\def\cbar  {\ensuremath{\overline c}\xspace}

\def\piz   {\ensuremath{\pi^0}\xspace}

\def\pip   {\ensuremath{\pi^+}\xspace}
\def\pim   {\ensuremath{\pi^-}\xspace}

\def\Kbar  {\kern 0.2em\overline{\kern -0.2em K}{}\xspace}

\def\Kz    {\ensuremath{K^0}\xspace}
\def\Kzb   {\ensuremath{\Kbar^0}\xspace}
\def\KzKzb {\ensuremath{\Kz \kern -0.16em \Kzb}\xspace}
\def\Kp    {\ensuremath{K^+}\xspace}
\def\Km    {\ensuremath{K^-}\xspace}

\def\KpKm  {\ensuremath{\Kp \kern -0.16em \Km}\xspace}
 
\def\KL    {\ensuremath{K^0_{\scriptscriptstyle\rm L}}\xspace}

\def\D       {\ensuremath{D}\xspace}
\def\Dbar    {\kern 0.2em\overline{\kern -0.2em D}{}\xspace}

\def\Dz      {\ensuremath{D^0}\xspace}
\def\Dzb     {\ensuremath{\Dbar^0}\xspace}
\def\DzDzb   {\ensuremath{\Dz {\kern -0.16em \Dzb}}\xspace}
\def\Dp      {\ensuremath{D^+}\xspace}
\def\Dm      {\ensuremath{D^-}\xspace}

\def\DpDm    {\ensuremath{\Dp {\kern -0.16em \Dm}}\xspace}

\def\Dstarp  {\ensuremath{D^{*+}}\xspace}
\def\Dstarm  {\ensuremath{D^{*-}}\xspace}

\def\Ds      {\ensuremath{D^+_s}\xspace}

\def\Bbar    {\kern 0.18em\overline{\kern -0.18em B}{}\xspace}

\def\Bz      {\ensuremath{B^0}\xspace}
\def\Bzb     {\ensuremath{\Bbar^0}\xspace}
\def\BzBzb   {\ensuremath{\Bz {\kern -0.16em \Bzb}}\xspace}
\def\Bu      {\ensuremath{B^+}\xspace}
\def\Bub     {\ensuremath{B^-}\xspace}

\def\BpBm    {\ensuremath{\Bu {\kern -0.16em \Bub}}\xspace}

\mathchardef\Upsilon="7107
\def\Y#1S{\ensuremath{\Upsilon{(#1S)}}\xspace}%

\mathchardef\Deltares="7101
\mathchardef\Xi="7104
\mathchardef\Lambda="7103
\mathchardef\Sigma="7106
\mathchardef\Omega="710A

\def\Deltabar{\kern 0.25em\overline{\kern -0.25em \Deltares}{}\xspace}
\def\Lbar{\kern 0.2em\overline{\kern -0.2em\Lambda\kern 0.05em}\kern-0.05em{}\xspace}
\def\Sigbar{\kern 0.2em\overline{\kern -0.2em \Sigma}{}\xspace}
\def\Xibar{\kern 0.2em\overline{\kern -0.2em \Xi}{}\xspace}
\def\Obar{\kern 0.2em\overline{\kern -0.2em \Omega}{}\xspace}
\def\Nbar{\kern 0.2em\overline{\kern -0.2em N}{}\xspace}
\def\Xb{\kern 0.2em\overline{\kern -0.2em X}{}\xspace}

\newcommand{\tev}{\ensuremath{\mathrm{\,Te\kern -0.1em V}}\xspace}
\newcommand{\gev}{\ensuremath{\mathrm{\,Ge\kern -0.1em V}}\xspace}
\newcommand{\mev}{\ensuremath{\mathrm{\,Me\kern -0.1em V}}\xspace}
\newcommand{\kev}{\ensuremath{\mathrm{\,ke\kern -0.1em V}}\xspace}
\newcommand{\ev}{\ensuremath{\mathrm{\,e\kern -0.1em V}}\xspace}
\newcommand{\gevc}{\ensuremath{{\mathrm{\,Ge\kern -0.1em V\!/}c}}\xspace}
\newcommand{\mevc}{\ensuremath{{\mathrm{\,Me\kern -0.1em V\!/}c}}\xspace}
\newcommand{\gevcc}{\ensuremath{{\mathrm{\,Ge\kern -0.1em V\!/}c^2}}\xspace}
\newcommand{\mevcc}{\ensuremath{{\mathrm{\,Me\kern -0.1em V\!/}c^2}}\xspace}

\def\cm   {\ensuremath{{\rm \,cm}}\xspace}

\def\mum  {\ensuremath{\,\upmu\nospaceunit{m}}\xspace}%

\def\invfb   {\ensuremath{\mbox{\,fb}^{-1}}\xspace}

\def\mus  {\ensuremath{\rm \,\mus}\xspace}

\def\ps   {\ensuremath{\rm \,ps}\xspace}
\def\fs   {\ensuremath{\rm \,fs}\xspace}

\def\mus        {\ensuremath{\,\mu{\rm s}}\xspace}    %
\def\ps         {\ensuremath{{\rm \,ps}}\xspace}  %

\def\to                 {\ensuremath{\rightarrow}\xspace}

\newcommand{\stat}{\ensuremath{\,\mathrm{(stat)}}\xspace}
\newcommand{\syst}{\ensuremath{\,\mathrm{(syst)}}\xspace}

\def\gsim{{~\raise.15em\hbox{$>$}\kern-.85em
          \lower.35em\hbox{$\sim$}~}\xspace}
\def\lsim{{~\raise.15em\hbox{$<$}\kern-.85em
          \lower.35em\hbox{$\sim$}~}\xspace}

\usepackage{ifthen} %

\newboolean{wordcount}
\setboolean{wordcount}{false} %

\ifthenelse{\boolean{wordcount}}%
{\usepackage{bibentry} %
 \usepackage{comment} %
 \excludecomment{align*} %
 \excludecomment{align} %
 \excludecomment{equation*} %
 \excludecomment{equation} %
 \excludecomment{eqnarray*} %
 \excludecomment{eqnarray} %
 \excludecomment{acknowledgments} %
 \def\maketitle{} %
}{
}

\makeatletter%
\begin{document}

\title{%
%
Precise measurement of the \Dz and \Dp lifetimes at \belletwo 
}
%
%
%
%
%
%
%
\newcommand{\instCPPM}{Aix Marseille Universit\'{e}, CNRS/IN2P3, CPPM, 13288 Marseille, France}
\newcommand{\instYerevan}{Alikhanyan National Science Laboratory, Yerevan 0036, Armenia}
\newcommand{\instBeihang}{Beihang University, Beijing 100191, China}
\newcommand{\instBNL}{Brookhaven National Laboratory, Upton, New York 11973, U.S.A.}
\newcommand{\instBINP}{Budker Institute of Nuclear Physics SB RAS, Novosibirsk 630090, Russian Federation}
\newcommand{\instCMU}{Carnegie Mellon University, Pittsburgh, Pennsylvania 15213, U.S.A.}
\newcommand{\instCinvestavIPN}{Centro de Investigacion y de Estudios Avanzados del Instituto Politecnico Nacional, Mexico City 07360, Mexico}
\newcommand{\instPrague}{Faculty of Mathematics and Physics, Charles University, 121 16 Prague, Czech Republic}
\newcommand{\instChiangMai}{Chiang Mai University, Chiang Mai 50202, Thailand}
\newcommand{\instChiba}{Chiba University, Chiba 263-8522, Japan}
\newcommand{\instChonnam}{Chonnam National University, Gwangju 61186, South Korea}
\newcommand{\instConacyt}{Consejo Nacional de Ciencia y Tecnolog\'{\i}a, Mexico City 03940, Mexico}
\newcommand{\instDESY}{Deutsches Elektronen--Synchrotron, 22607 Hamburg, Germany}
\newcommand{\instDuke}{Duke University, Durham, North Carolina 27708, U.S.A.}
\newcommand{\instITAR}{Institute of Theoretical and Applied Research (ITAR), Duy Tan University, Hanoi 100000, Vietnam}
\newcommand{\instRomaENEA}{ENEA Casaccia, I-00123 Roma, Italy}
\newcommand{\instFuJen}{Department of Physics, Fu Jen Catholic University, Taipei 24205, Taiwan}
\newcommand{\instFudan}{Key Laboratory of Nuclear Physics and Ion-beam Application (MOE) and Institute of Modern Physics, Fudan University, Shanghai 200443, China}
\newcommand{\instGoettingen}{II. Physikalisches Institut, Georg-August-Universit\"{a}t G\"{o}ttingen, 37073 G\"{o}ttingen, Germany}
\newcommand{\instGifu}{Gifu University, Gifu 501-1193, Japan}
\newcommand{\instSOKENDAI}{The Graduate University for Advanced Studies (SOKENDAI), Hayama 240-0193, Japan}
\newcommand{\instGyeongsang}{Gyeongsang National University, Jinju 52828, South Korea}
\newcommand{\instHanyang}{Department of Physics and Institute of Natural Sciences, Hanyang University, Seoul 04763, South Korea}
\newcommand{\instKEK}{High Energy Accelerator Research Organization (KEK), Tsukuba 305-0801, Japan}
\newcommand{\instJPARC}{J-PARC Branch, KEK Theory Center, High Energy Accelerator Research Organization (KEK), Tsukuba 305-0801, Japan}
\newcommand{\instHiroshima}{Hiroshima University, Higashi-Hiroshima, Hiroshima 739-8530, Japan}
\newcommand{\instFrascati}{INFN Laboratori Nazionali di Frascati, I-00044 Frascati, Italy}
\newcommand{\instNapoliINFN}{INFN Sezione di Napoli, I-80126 Napoli, Italy}
\newcommand{\instPadovaINFN}{INFN Sezione di Padova, I-35131 Padova, Italy}
\newcommand{\instPerugiaINFN}{INFN Sezione di Perugia, I-06123 Perugia, Italy}
\newcommand{\instPisaINFN}{INFN Sezione di Pisa, I-56127 Pisa, Italy}
\newcommand{\instRomaINFN}{INFN Sezione di Roma, I-00185 Roma, Italy}
\newcommand{\instRomaTreINFN}{INFN Sezione di Roma Tre, I-00146 Roma, Italy}
\newcommand{\instTorinoINFN}{INFN Sezione di Torino, I-10125 Torino, Italy}
\newcommand{\instTriesteINFN}{INFN Sezione di Trieste, I-34127 Trieste, Italy}
\newcommand{\instIISER}{Indian Institute of Science Education and Research Mohali, SAS Nagar, 140306, India}
\newcommand{\instIITBhubaneswar}{Indian Institute of Technology Bhubaneswar, Satya Nagar 751007, India}
\newcommand{\instIITGuwahati}{Indian Institute of Technology Guwahati, Assam 781039, India}
\newcommand{\instIITHyderabad}{Indian Institute of Technology Hyderabad, Telangana 502285, India}
\newcommand{\instIITMadras}{Indian Institute of Technology Madras, Chennai 600036, India}
\newcommand{\instIndiana}{Indiana University, Bloomington, Indiana 47408, U.S.A.}
\newcommand{\instIHEPRussia}{Institute for High Energy Physics, Protvino 142281, Russian Federation}
\newcommand{\instHEPHYVienna}{Institute of High Energy Physics, Vienna 1050, Austria}
\newcommand{\instIHEPChina}{Institute of High Energy Physics, Chinese Academy of Sciences, Beijing 100049, China}
\newcommand{\instIPP}{Institute of Particle Physics (Canada), Victoria, British Columbia V8W 2Y2, Canada}
\newcommand{\instIOP}{Institute of Physics, Vietnam Academy of Science and Technology (VAST), Hanoi, Vietnam}
\newcommand{\instIFIC}{Instituto de Fisica Corpuscular, Paterna 46980, Spain}
\newcommand{\instISU}{Iowa State University, Ames, Iowa 50011, U.S.A.}
\newcommand{\instJAEA}{Advanced Science Research Center, Japan Atomic Energy Agency, Naka 319-1195, Japan}
\newcommand{\instMainz}{Institut f\"{u}r Kernphysik, Johannes Gutenberg-Universit\"{a}t Mainz, D-55099 Mainz, Germany}
\newcommand{\instGiessen}{Justus-Liebig-Universit\"{a}t Gie\ss{}en, 35392 Gie\ss{}en, Germany}
\newcommand{\instKarlsruhe}{Institut f\"{u}r Experimentelle Teilchenphysik, Karlsruher Institut f\"{u}r Technologie, 76131 Karlsruhe, Germany}
\newcommand{\instKitasato}{Kitasato University, Sagamihara 252-0373, Japan}
\newcommand{\instKISTI}{Korea Institute of Science and Technology Information, Daejeon 34141, South Korea}
\newcommand{\instKoreaUnivKU}{Korea University, Seoul 02841, South Korea}
\newcommand{\instKSU}{Kyoto Sangyo University, Kyoto 603-8555, Japan}
\newcommand{\instKyungpook}{Kyungpook National University, Daegu 41566, South Korea}
\newcommand{\instLPI}{P.N. Lebedev Physical Institute of the Russian Academy of Sciences, Moscow 119991, Russian Federation}
\newcommand{\instLNNU}{Liaoning Normal University, Dalian 116029, China}
\newcommand{\instLMU}{Ludwig Maximilians University, 80539 Munich, Germany}
\newcommand{\instLuther}{Luther College, Decorah, Iowa 52101, U.S.A.}
\newcommand{\instMNITJaipur}{Malaviya National Institute of Technology Jaipur, Jaipur 302017, India}
\newcommand{\instMPP}{Max-Planck-Institut f\"{u}r Physik, 80805 M\"{u}nchen, Germany}
\newcommand{\instMPGHLL}{Semiconductor Laboratory of the Max Planck Society, 81739 M\"{u}nchen, Germany}
\newcommand{\instMcGill}{McGill University, Montr\'{e}al, Qu\'{e}bec, H3A 2T8, Canada}
\newcommand{\instMEPhI}{Moscow Physical Engineering Institute, Moscow 115409, Russian Federation}
\newcommand{\instNagoya}{Graduate School of Science, Nagoya University, Nagoya 464-8602, Japan}
\newcommand{\instNagoyaIAR}{Institute for Advanced Research, Nagoya University, Nagoya 464-8602, Japan}
\newcommand{\instNagoyaKMI}{Kobayashi-Maskawa Institute, Nagoya University, Nagoya 464-8602, Japan}
\newcommand{\instNaraWu}{Nara Women's University, Nara 630-8506, Japan}
\newcommand{\instHSE}{National Research University Higher School of Economics, Moscow 101000, Russian Federation}
\newcommand{\instNTUTaiwan}{Department of Physics, National Taiwan University, Taipei 10617, Taiwan}
\newcommand{\instNUUTaiwan}{National United University, Miao Li 36003, Taiwan}
\newcommand{\instKrakow}{H. Niewodniczanski Institute of Nuclear Physics, Krakow 31-342, Poland}
\newcommand{\instNiigata}{Niigata University, Niigata 950-2181, Japan}
\newcommand{\instNSU}{Novosibirsk State University, Novosibirsk 630090, Russian Federation}
\newcommand{\instOkinawa}{Okinawa Institute of Science and Technology, Okinawa 904-0495, Japan}
\newcommand{\instOsakaCity}{Osaka City University, Osaka 558-8585, Japan}
\newcommand{\instRCNP}{Research Center for Nuclear Physics, Osaka University, Osaka 567-0047, Japan}
\newcommand{\instPNNL}{Pacific Northwest National Laboratory, Richland, Washington 99352, U.S.A.}
\newcommand{\instPanjab}{Panjab University, Chandigarh 160014, India}
\newcommand{\instPanjabPAU}{Punjab Agricultural University, Ludhiana 141004, India}
\newcommand{\instRIKENMSL}{Meson Science Laboratory, Cluster for Pioneering Research, RIKEN, Saitama 351-0198, Japan}
\newcommand{\instXavier}{St. Francis Xavier University, Antigonish, Nova Scotia, B2G 2W5, Canada}
\newcommand{\instSeoul}{Seoul National University, Seoul 08826, South Korea}
\newcommand{\instSPU}{Showa Pharmaceutical University, Tokyo 194-8543, Japan}
\newcommand{\instSoochow}{Soochow University, Suzhou 215006, China}
\newcommand{\instSoongsil}{Soongsil University, Seoul 06978, South Korea}
\newcommand{\instLjubljanaJSI}{J. Stefan Institute, 1000 Ljubljana, Slovenia}
\newcommand{\instKyiv}{Taras Shevchenko National Univ. of Kiev, Kiev, Ukraine}
\newcommand{\instTata}{Tata Institute of Fundamental Research, Mumbai 400005, India}
\newcommand{\instTUM}{Department of Physics, Technische Universit\"{a}t M\"{u}nchen, 85748 Garching, Germany}
\newcommand{\instTelAviv}{Tel Aviv University, School of Physics and Astronomy, Tel Aviv, 69978, Israel}
\newcommand{\instToho}{Toho University, Funabashi 274-8510, Japan}
\newcommand{\instTohoku}{Department of Physics, Tohoku University, Sendai 980-8578, Japan}
\newcommand{\instTitech}{Tokyo Institute of Technology, Tokyo 152-8550, Japan}
\newcommand{\instTokyoMetropolitan}{Tokyo Metropolitan University, Tokyo 192-0397, Japan}
\newcommand{\instUAS}{Universidad Autonoma de Sinaloa, Sinaloa 80000, Mexico}
\newcommand{\instNapoliUNIV}{Dipartimento di Scienze Fisiche, Universit\`{a} di Napoli Federico II, I-80126 Napoli, Italy}
\newcommand{\instPadovaUNIV}{Dipartimento di Fisica e Astronomia, Universit\`{a} di Padova, I-35131 Padova, Italy}
\newcommand{\instPerugiaUNIV}{Dipartimento di Fisica, Universit\`{a} di Perugia, I-06123 Perugia, Italy}
\newcommand{\instPisaUNIV}{Dipartimento di Fisica, Universit\`{a} di Pisa, I-56127 Pisa, Italy}
\newcommand{\instRomaTreUNIV}{Dipartimento di Matematica e Fisica, Universit\`{a} di Roma Tre, I-00146 Roma, Italy}
\newcommand{\instTorinoUNIV}{Dipartimento di Fisica, Universit\`{a} di Torino, I-10125 Torino, Italy}
\newcommand{\instTriesteUNIV}{Dipartimento di Fisica, Universit\`{a} di Trieste, I-34127 Trieste, Italy}
\newcommand{\instMontreal}{Universit\'{e} de Montr\'{e}al, Physique des Particules, Montr\'{e}al, Qu\'{e}bec, H3C 3J7, Canada}
\newcommand{\instIJCLab}{Universit\'{e} Paris-Saclay, CNRS/IN2P3, IJCLab, 91405 Orsay, France}
\newcommand{\instIPHC}{Universit\'{e} de Strasbourg, CNRS, IPHC, UMR 7178, 67037 Strasbourg, France}
\newcommand{\instAdelaide}{Department of Physics, University of Adelaide, Adelaide, South Australia 5005, Australia}
\newcommand{\instBonn}{University of Bonn, 53115 Bonn, Germany}
\newcommand{\instUBC}{University of British Columbia, Vancouver, British Columbia, V6T 1Z1, Canada}
\newcommand{\instCincinnati}{University of Cincinnati, Cincinnati, Ohio 45221, U.S.A.}
\newcommand{\instFlorida}{University of Florida, Gainesville, Florida 32611, U.S.A.}
\newcommand{\instHawaii}{University of Hawaii, Honolulu, Hawaii 96822, U.S.A.}
\newcommand{\instHeidelberg}{University of Heidelberg, 68131 Mannheim, Germany}
\newcommand{\instLjubljanaUniLJ}{Faculty of Mathematics and Physics, University of Ljubljana, 1000 Ljubljana, Slovenia}
\newcommand{\instLouisville}{University of Louisville, Louisville, Kentucky 40292, U.S.A.}
\newcommand{\instMalaya}{National Centre for Particle Physics, University Malaya, 50603 Kuala Lumpur, Malaysia}
\newcommand{\instLjubljanaUM}{Faculty of Chemistry and Chemical Engineering, University of Maribor, 2000 Maribor, Slovenia}
\newcommand{\instMelbourne}{School of Physics, University of Melbourne, Victoria 3010, Australia}
\newcommand{\instMississippi}{University of Mississippi, University, Mississippi 38677, U.S.A.}
\newcommand{\instUOM}{University of Miyazaki, Miyazaki 889-2192, Japan}
\newcommand{\instPittsburgh}{University of Pittsburgh, Pittsburgh, Pennsylvania 15260, U.S.A.}
\newcommand{\instUSTC}{University of Science and Technology of China, Hefei 230026, China}
\newcommand{\instSAlabama}{University of South Alabama, Mobile, Alabama 36688, U.S.A.}
\newcommand{\instSCarolina}{University of South Carolina, Columbia, South Carolina 29208, U.S.A.}
\newcommand{\instSydney}{School of Physics, University of Sydney, New South Wales 2006, Australia}
\newcommand{\instUTokyo}{Department of Physics, University of Tokyo, Tokyo 113-0033, Japan}
\newcommand{\instEri}{Earthquake Research Institute, University of Tokyo, Tokyo 113-0032, Japan}
\newcommand{\instIPMU}{Kavli Institute for the Physics and Mathematics of the Universe (WPI), University of Tokyo, Kashiwa 277-8583, Japan}
\newcommand{\instVictoria}{University of Victoria, Victoria, British Columbia, V8W 3P6, Canada}
\newcommand{\instVPI}{Virginia Polytechnic Institute and State University, Blacksburg, Virginia 24061, U.S.A.}
\newcommand{\instWayneState}{Wayne State University, Detroit, Michigan 48202, U.S.A.}
\newcommand{\instYamagata}{Yamagata University, Yamagata 990-8560, Japan}
\newcommand{\instYonsei}{Yonsei University, Seoul 03722, South Korea}
\affiliation{\instCPPM}
\affiliation{\instYerevan}
\affiliation{\instBNL}
\affiliation{\instBINP}
\affiliation{\instCMU}
\affiliation{\instCinvestavIPN}
\affiliation{\instPrague}
\affiliation{\instChiangMai}
\affiliation{\instChiba}
\affiliation{\instConacyt}
\affiliation{\instDESY}
\affiliation{\instDuke}
\affiliation{\instITAR}
\affiliation{\instRomaENEA}
\affiliation{\instFuJen}
\affiliation{\instFudan}
\affiliation{\instGifu}
\affiliation{\instSOKENDAI}
\affiliation{\instGyeongsang}
\affiliation{\instHanyang}
\affiliation{\instKEK}
\affiliation{\instJPARC}
\affiliation{\instHiroshima}
\affiliation{\instFrascati}
\affiliation{\instNapoliINFN}
\affiliation{\instPadovaINFN}
\affiliation{\instPerugiaINFN}
\affiliation{\instPisaINFN}
\affiliation{\instRomaTreINFN}
\affiliation{\instTorinoINFN}
\affiliation{\instTriesteINFN}
\affiliation{\instIISER}
\affiliation{\instIITBhubaneswar}
\affiliation{\instIITHyderabad}
\affiliation{\instIITMadras}
\affiliation{\instIndiana}
\affiliation{\instIHEPRussia}
\affiliation{\instHEPHYVienna}
\affiliation{\instIPP}
\affiliation{\instIOP}
\affiliation{\instIFIC}
\affiliation{\instISU}
\affiliation{\instJAEA}
\affiliation{\instMainz}
\affiliation{\instGiessen}
\affiliation{\instKarlsruhe}
\affiliation{\instKitasato}
\affiliation{\instKISTI}
\affiliation{\instKoreaUnivKU}
\affiliation{\instKSU}
\affiliation{\instKyungpook}
\affiliation{\instLPI}
\affiliation{\instLNNU}
\affiliation{\instLMU}
\affiliation{\instLuther}
\affiliation{\instMNITJaipur}
\affiliation{\instMPP}
\affiliation{\instMcGill}
\affiliation{\instMEPhI}
\affiliation{\instNagoya}
\affiliation{\instNagoyaIAR}
\affiliation{\instNagoyaKMI}
\affiliation{\instNaraWu}
\affiliation{\instHSE}
\affiliation{\instNTUTaiwan}
\affiliation{\instKrakow}
\affiliation{\instNiigata}
\affiliation{\instNSU}
\affiliation{\instOkinawa}
\affiliation{\instOsakaCity}
\affiliation{\instRCNP}
\affiliation{\instPNNL}
\affiliation{\instPanjab}
\affiliation{\instPanjabPAU}
\affiliation{\instRIKENMSL}
\affiliation{\instXavier}
\affiliation{\instSPU}
\affiliation{\instSoochow}
\affiliation{\instSoongsil}
\affiliation{\instLjubljanaJSI}
\affiliation{\instKyiv}
\affiliation{\instTata}
\affiliation{\instTUM}
\affiliation{\instTelAviv}
\affiliation{\instToho}
\affiliation{\instTokyoMetropolitan}
\affiliation{\instUAS}
\affiliation{\instNapoliUNIV}
\affiliation{\instPadovaUNIV}
\affiliation{\instPerugiaUNIV}
\affiliation{\instPisaUNIV}
\affiliation{\instRomaTreUNIV}
\affiliation{\instTorinoUNIV}
\affiliation{\instTriesteUNIV}
\affiliation{\instIJCLab}
\affiliation{\instIPHC}
\affiliation{\instBonn}
\affiliation{\instUBC}
\affiliation{\instCincinnati}
\affiliation{\instFlorida}
\affiliation{\instHawaii}
\affiliation{\instLjubljanaUniLJ}
\affiliation{\instLouisville}
\affiliation{\instLjubljanaUM}
\affiliation{\instMelbourne}
\affiliation{\instMississippi}
\affiliation{\instUOM}
\affiliation{\instPittsburgh}
\affiliation{\instUSTC}
\affiliation{\instSAlabama}
\affiliation{\instSydney}
\affiliation{\instUTokyo}
\affiliation{\instIPMU}
\affiliation{\instVictoria}
\affiliation{\instVPI}
\affiliation{\instWayneState}
\affiliation{\instYamagata}
\affiliation{\instYonsei}
  \author{F.~Abudin{\'e}n}\affiliation{\instTriesteINFN} %
  \author{I.~Adachi}\affiliation{\instKEK}\affiliation{\instSOKENDAI} %
  \author{K.~Adamczyk}\affiliation{\instKrakow} %
  \author{L.~Aggarwal}\affiliation{\instPanjab} %
  \author{H.~Ahmed}\affiliation{\instXavier} %
  \author{H.~Aihara}\affiliation{\instUTokyo} %
  \author{N.~Akopov}\affiliation{\instYerevan} %
  \author{A.~Aloisio}\affiliation{\instNapoliUNIV}\affiliation{\instNapoliINFN} %
  \author{N.~Anh~Ky}\affiliation{\instIOP}\affiliation{\instITAR} %
  \author{D.~M.~Asner}\affiliation{\instBNL} %
% ...
%
\collaboration{Belle II Collaboration}
 \begin{abstract}
%
%
We report a measurement of the \Dz and \Dp lifetimes using \DzToKpi and \DpToKpipi decays reconstructed in $\epem\to\c\cbar$ data recorded by the \belletwo experiment at the SuperKEKB asymmetric-energy \epem collider. The data, collected at center-of-mass energies at or near the $\Upsilon(4S)$ resonance, correspond to an integrated luminosity of \lumi. The results, $\tau(\Dz) = \tauDz\pm\tauDzStat\stat\pm\tauDzSyst\syst\fs$ and $\tau(\Dp) = \tauDp\pm\tauDpStat\stat\pm\tauDpSyst\syst\fs$, are the most precise to date and are consistent with previous determinations. 
 \end{abstract}

\maketitle

%
Accurate predictions of lifetimes of weakly decaying charmed and bottom hadrons are challenging because they involve strong-interaction theory at low energy. Predictions must resort to effective models, such as the heavy-quark expansion~\cite{Neubert:1997gu,Uraltsev:2000qw,Lenz:2013aua,Lenz:2014jha,Kirk:2017juj,Cheng:2018rkz}, which also underpin strong-interaction calculations required for the determination of fundamental standard-model parameters from hadron-decay measurements (\eg, to extract the strength of quark-mixing couplings from decay widths). Precise lifetime measurements provide excellent tests of such effective models. Lifetimes are also important inputs for a wide variety of studies because they are needed to compare measured decay branching fractions to predictions for partial decay widths.

Weakly decaying charmed hadrons have lifetimes ranging from about 0.1 to 1\ps~\cite{pdg}. The world averages of the \Dz and \Dp lifetimes, $410.1\pm1.5\fs$ and $1040\pm7\fs$, are almost exclusively determined from systematically limited per-mille-precision measurements made by FOCUS two decades ago~\cite{pdg,Link:2002bx}. Recently, the LHCb collaboration precisely measured the lifetimes of the \Ds meson and charmed baryons relative to that of the \Dp meson~\cite{Aaij:2017vqj,Aaij:2018dso,Aaij:2018wzf,Aaij:2019lwg}. Such relative measurements minimize systematic uncertainties due to decay-time-biasing event-selection criteria that are particularly severe at hadron colliders. By contrast, experiments at \epem colliders, owing to the reconstruction of large charmed hadron yields without decay-time-biasing selections, have a great potential for absolute lifetime measurements. With the first layer of its vertex detector only 1.4\cm away from the interaction region, the \belletwo experiment at the SuperKEKB asymmetric-energy \epem collider~\cite{Abe:2010gxa,Akai:2018mbz} obtains a decay-time resolution two times better than the Belle and \babar experiments~\cite{Bevan:2014iga}, enabling high precision for the measurement of charmed lifetimes with early data. To limit systematic uncertainties this potential must be complemented with an accurate vertex-detector alignment, a precise calibration of final-state particle momenta, and powerful background discrimination.

In this Letter, we report high-precision measurements of the \Dz and \Dp lifetimes using $\Dstarp\to\Dz(\to\Km\pip)\pip$ and $\Dstarp\to\Dp(\to\Km\pip\pip)\piz$ decays reconstructed in the data collected by \belletwo during 2019 and the first half of 2020 at center-of-mass energies at or near the $\Upsilon(4S)$ resonance. (Charge-conjugate decays are implied throughout.) The data correspond to an integrated luminosity of \lumi. At \belletwo, \Dstarp mesons from $\epem\to\c\cbar$ events are produced with boosts that displace the \Dz and \Dp decay points from those of production by approximately 200\mum and 500\mum on average, respectively. The decay time is measured from this displacement, $\vec{L}$, projected onto the direction of the momentum, $\vec{p}$, as $t = m_D\vec{L}\cdot\vec{p}/|\vec{p}|^2$, where $m_D$ is the known mass of the relevant \D meson~\cite{pdg}. The decay-time uncertainty, \sigmat, is calculated by propagating the uncertainties in $\vec{L}$ and $\vec{p}$, including their correlations. The lifetimes are measured using a fit to the $(t,\sigmat)$ distributions of the reconstructed decay candidates. The sample selection and fit strategy have been optimized and validated using simulation; however, no input from simulation is used in the fit to data. To avoid bias, we inspected the lifetimes measured with the full data sample only after the entire analysis procedure was finalized and all uncertainties were determined. However, we examined the results from the subset of data collected during 2019 (approximately 13\% of the total) before the analysis was complete.

The \belletwo detector~\cite{Abe:2010gxa} consists of several subsystems arranged in a cylindrical structure around the beam pipe. The tracking system consists of a two-layer silicon-pixel detector (PXD), surrounded by a four-layer double-sided silicon-strip detector (SVD) and a 56-layer central drift chamber (CDC). Only two out of 12 ladders were installed in the second layer of the PXD for this data sample. The combined PXD and SVD system provides average decay-time resolutions of about 70\fs and 60\fs, respectively, for the \Dz and \Dp decays considered here. A time-of-propagation counter and an aerogel ring-imaging Cherenkov counter that cover the barrel and forward end cap regions of the detector, respectively, are essential for charged-particle identification. The electromagnetic calorimeter fills the remaining volume inside a $1.5\,\rm{T}$ superconducting solenoid and serves to reconstruct photons and electrons. A dedicated system to identify \KL mesons and muons is installed in the outermost part of the detector. The $z$ axis of the laboratory frame is defined as the central axis of the solenoid, with its positive direction determined by the direction of the electron beam.

The simulation uses \texttt{KKMC}~\cite{Jadach:1999vf} to generate quark-antiquark pairs from \epem collisions, \texttt{PYTHIA8}~\cite{Sjostrand:2014zea} for hadronization, \texttt{EVTGEN}~\cite{Lange:2001uf} for the decay of the generated hadrons, and \texttt{GEANT4}~\cite{Agostinelli:2002hh} for the detector response.

The reconstruction~\cite{Kuhr:2018lps,BelleIITrackingGroup:2020hpx,Simon:1960} and selection of the signal candidates avoid any requirement that could bias the decay time or introduce a variation of the efficiency as a function of decay time, as checked in simulation. Events are first selected by vetoing events consistent with Bhabha scattering and by requiring at least three tracks with loose upper bounds on their impact parameters and with transverse momenta greater than 200\mevc. These three tracks are not necessarily associated with the decay modes being reconstructed.

Candidate \DzToKpi decays are formed using pairs of oppositely charged tracks. Each track must have a hit in the first layer of the PXD, at least one hit in the SVD, at least 20 hits in the CDC, and be identified as a kaon, if negative, or else a pion. Low-momentum pion candidates are tracks consistent with originating from the interaction region that are required to have hits both in the SVD and CDC. They are combined with \Dz candidates to form $\Dstarp\to\Dz\pip$ decays. A global decay-chain vertex fit~\cite{treefitter} constrains the tracks according to the decay topology and constrains the \Dstarp candidate to originate from the measured position of the \epem interaction region (IR). Only candidates with fit $\chi^2$ probabilities larger than $0.01$ are retained for further analysis. The IR has typical dimensions of 250\mum along the $z$ axis and of 10\mum and 0.3\mum in the two directions of the transverse plane. Its position and size vary over data-taking and are regularly measured using $\epem\to\mup\mun$ events. The mass of the \Dz candidate, \mDKpi, must be in the range $[1.75,2.00]\gevcc$. The difference between the \Dstarp- and \Dz-candidate masses, \dm, must satisfy $144.94<\dm<145.90\mevcc$ ($\pm3$ times the \dm resolution around the signal peak). Since the \Dz is assumed to originate from the IR, charmed mesons originating from displaced decays of bottom mesons would bias the lifetime measurement. They are suppressed to a negligible rate by requiring that the momentum of the \Dstarp in the \epem center-of-mass system exceeds $2.5\gevc$. After requiring $1.851<\mDKpi<1.878\gevcc$ (signal region), multiple \Dstarp candidates occur in a few per mille of the selected events. In such events, one randomly selected candidate is retained for subsequent analysis.

\begin{figure}[t!]
\centering
\includegraphics[width=\linewidth]{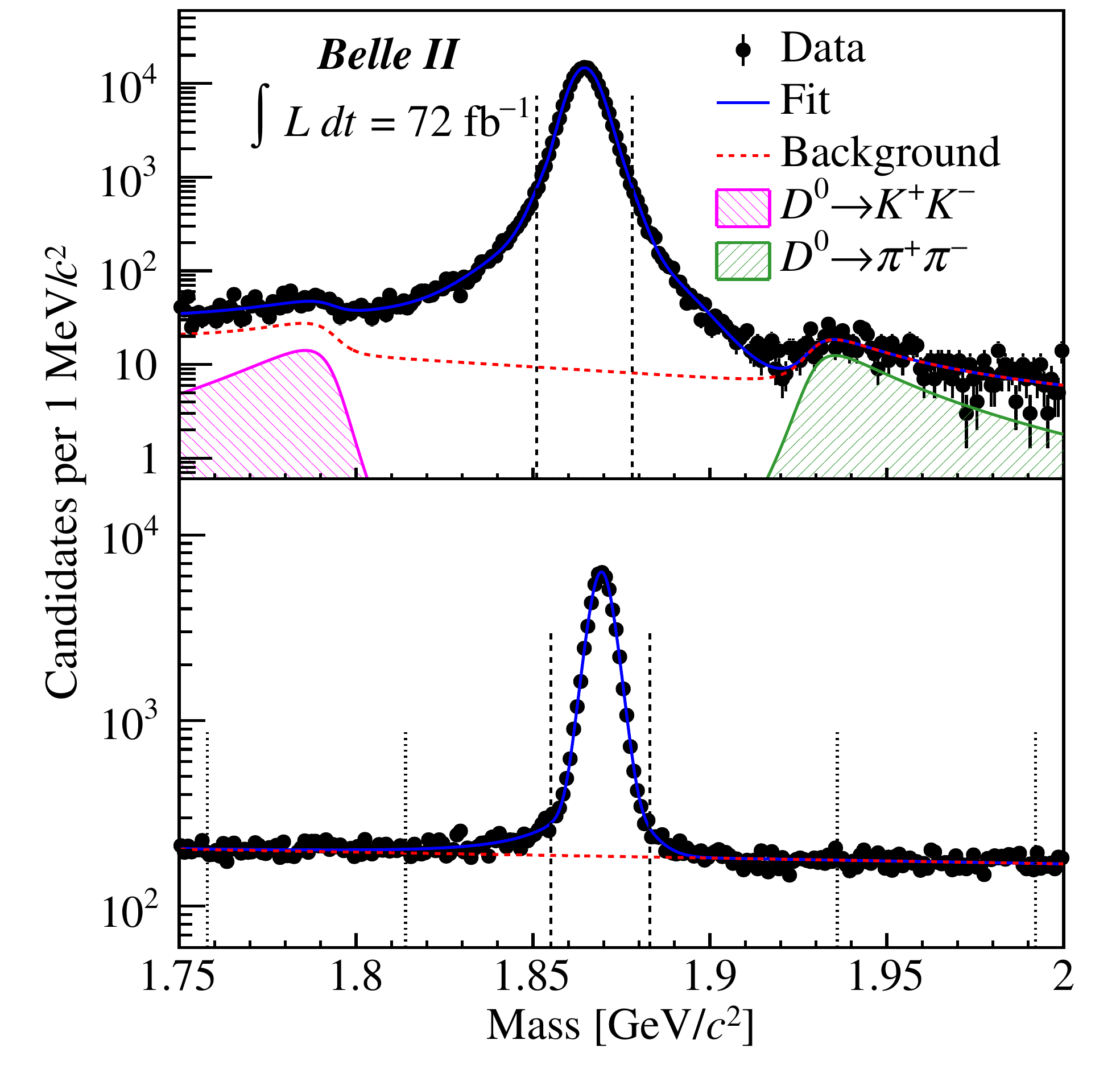}\\
\caption{Mass distributions of (top) \DzToKpi and (bottom) \DpToKpipi candidates with fit projections overlaid. The vertical dashed and (for the bottom plot) dotted lines indicate the signal regions and the sideband, respectively.\label{fig:massfit}}
\end{figure}

The signal region contains approximately $171\times10^3$ candidates with a signal purity of about 99.8\%, as determined from a binned least-squares fit to the \mDKpi distribution (Fig.~\ref{fig:massfit}). In the fit, the \DzToKpi signal is modeled with the sum of two Gaussian distributions and a Crystal Ball function~\cite{Skwarnicki:1986xj}; misidentified decays of $\Dz\to\pip\pim$ and $\Dz\to\Kp\Km$, each modeled with a Johnson's $S_U$ distribution~\cite{johnson} with parameters determined from simulation, do not enter the signal region; the remaining background, modeled with an exponential distribution, is dominated by candidates formed by random combinations of particles.

The selection of the $\Dstarp\to\Dp(\to\Km\pip\pip)\piz$ candidates follows similar criteria to those for the \Dz mode, but with more stringent requirements to suppress a larger background contamination. Tracks identified as kaons or pions are required to have a hit in the first layer of the PXD, at least one hit in the SVD, and at least 30 hits in the CDC. They are combined to form \DpToKpipi candidates. To suppress backgrounds from misreconstructed charmed-hadron decays, such as four-body hadronic or semileptonic decays, the lower-momentum pion must have momentum exceeding 350\mevc and the higher-momentum pion must not be identified as a lepton. Candidate $\piz\to\gamma\gamma$ decays are reconstructed using photon candidates from calorimetric energy clusters that are not associated with a track. Each photon energy must be larger than 80, 30, or 60\mev if detected in the forward, central, or backward region, respectively, of the calorimeter. Neutral-pion candidates with masses in the range $[120,145]\mevcc$ and momenta larger than 150\mevc are combined with \Dp candidates to form $\Dstarp\to\Dp\piz$ decays. The \Dstarp decay chain is fit using IR and \piz-mass constraints. Only candidates with fit $\chi^2$ probabilities larger than $0.01$ are retained. The mass of the \Dp candidate, \mDp, must be in the range $[1.75,2.00]\gevcc$ and the difference between the \Dstarp and \Dp masses in the range $[138,143]\mevcc$ ($\pm3$ times the \dm resolution around the signal peak). The momentum of the \Dstarp in the \epem center-of-mass system must exceed $2.6\gevc$ to suppress \Dstarp candidates from bottom mesons. This requirement is tighter than that used for \Dz candidates because of the less-precise \piz-momentum resolution.

The signal region in \mDp is defined as $[1.855,1.883]\gevcc$ (Fig.~\ref{fig:massfit}). It contains approximately $59\times10^3$ candidates after randomly selecting one \Dstarp candidate for the percent-level fraction of events where more than one is found. A binned least-squares fit to the \mDp distribution identifies about 9\% of candidates in the signal region as background. Simulation shows that such background is composed of misreconstructed charmed decays and random track combinations. In the fit, the \DpToKpipi signal is modeled with the sum of two Gaussian distributions and a Crystal Ball function; the background is modeled with an exponential distribution.

The lifetimes are determined with unbinned maximum-likelihood fits to the $(t,\sigmat)$ distributions of the candidates populating the signal regions. Each signal probability-density function (PDF) is the convolution of an exponential distribution in $t$ with a resolution function that depends on \sigmat, multiplied by the PDF of \sigmat. In the \Dp case, simulation shows that a Gaussian distribution is sufficient to model the resolution function. The mean of the resolution function is allowed to float in the fit to account for a possible bias in the determination of the decay time; the width is the per-candidate \sigmat scaled by a free parameter $s$ to account for a possible misestimation of the decay-time uncertainty. The fit returns $s\approx1.12$ ($1.29$) for the \Dz (\Dp) sample. In the \Dz case, an additional Gaussian distribution is needed to describe the 3\% of candidates with poorer resolution. This second component shares its mean with the principal component but has its own free scaling parameter ($s'\approx2.5$) for the broader width.

In the \Dz case, the signal region contains a 0.2\% fraction of background candidates. Sensitivity to the background contamination and its effects on the decay-time distribution is very limited. For the sake of simplicity, the background is neglected in the fit and a systematic uncertainty is later assigned. In the \Dp case, the signal region contains a non-negligible amount of background, which is accounted for in the fit. The background is modeled using data with \mDp in the \emph{sideband} $[1.758,1.814]\cup[1.936,1.992]\gevcc$ (Fig.~\ref{fig:massfit}), which is assumed to contain exclusively background candidates and be representative of the background in the signal region, as verified in simulation. The background PDF consists of a zero-lifetime component and two exponential components, all convolved with a Gaussian resolution function having a free mean and a width corresponding to $s\sigma_t$. To better constrain the background parameters, a simultaneous fit to the candidates in the signal region and sideband is performed. The background fraction is Gaussian constrained in the fit to $(8.78\pm0.05)\%$, as measured in the \mDp fit. 

The PDF of \sigmat is a histogram template derived directly from the data. In the fit to the \Dz sample, the template is derived assuming that all candidates in the signal region are signal decays. In the fit to the \Dp sample, the template is derived from the candidates in the signal region by subtracting the scaled distribution of the sideband data. The PDF of \sigmat for the background is obtained directly from the sideband data.

The lifetime fits are tested on fully simulated data and on sets of data generated by randomly sampling the PDF with parameters fixed to the values found in the fits to the data. All tests yield unbiased results and expected parameter uncertainties, independent of the assumed values of the \Dz and \Dp lifetimes.

\begin{figure}[t!]
\centering
\includegraphics[width=\linewidth]{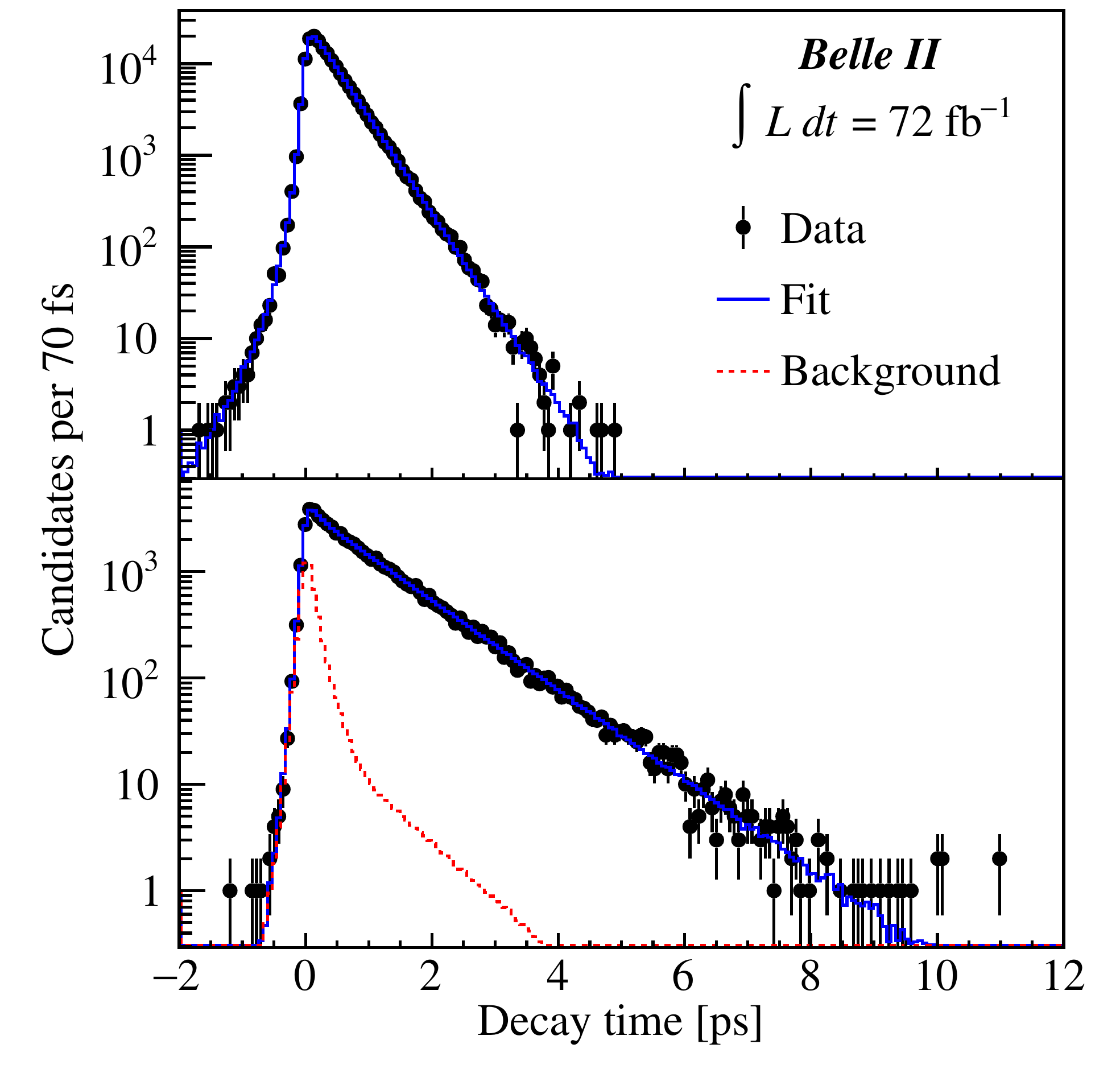}\\
\caption{Decay-time distributions of (top) \DzToKpi and (bottom) \DpToKpipi candidates in their respective signal regions with fit projections overlaid.\label{fig:lifetime-fit}}
\end{figure}

The decay-time distributions of the data, with fit projections overlaid, are shown in Fig.~\ref{fig:lifetime-fit}. The measured \Dz and \Dp lifetimes $\tauDzKpi\pm\tauDzStatKpi\stat\pm\tauDzSystKpi\syst\fs$ and 
$\tauDp\pm\tauDpStat\stat\pm\tauDpSyst\syst\fs$, respectively, are consistent with their world averages~\cite{pdg}. The systematic uncertainties arise from the sources listed in Table~\ref{tab:syst} and described below. The total systematic uncertainty is the sum in quadrature of the individual components.

\begin{table}[t]
\centering
\caption{Systematic uncertainties.\label{tab:syst}}
\begin{tabular}{lcc}
\hline
Source & $\tau(\Dz)$ [fs] & $\tau(\Dp)$ [fs]\\
\hline
Resolution model   & 0.16 & 0.39 \\
Backgrounds        & 0.24 & 2.52 \\
Detector alignment & 0.72 & 1.70 \\
Momentum scale     & 0.19 & 0.48 \\
\hline
Total  & \tauDzSystKpi0 & \tauDpSyst0 \\
\hline
\end{tabular}
\end{table}

The decay time and decay-time uncertainty are observed to be correlated in data and simulation reproduces these effects well. The dominant effect is that small \sigmat values correspond to larger true decay times (and \textit{vice versa}). These correlations, when neglected in the fits, result in an imperfect description of the $t$ distribution as a function of \sigmat. To quantify the impact on the results, our model that neglects the correlations is fit to 1000 samples of signal-only simulated decays, each the same size as the data. The samples are obtained by resampling, with repetition, a set of simulated \epem collisions corresponding to an integrated luminosity of 500\invfb. Upper bounds of 0.16\fs and 0.39\fs on the average absolute deviations of the measured lifetimes from their true values are derived and assigned as the systematic uncertainty due to the imperfect resolution model for the \DzToKpi and \DpToKpipi cases, respectively. For signal decays, the bias of the decay-time resolution function depends nearly linearly on the candidate mass and may not average out when the mass range is restricted. Varying the boundaries of the signal region shows that such a correlation has a negligible effect upon the measured lifetimes.

The background neglected in the \DzToKpi fit could result in a systematic bias on the measured lifetime. To estimate the size of the bias, we fit our model that neglects the background to 500 resampled sets of simulated \epem collisions, each having the same size and signal-to-background proportion as the data. The measured lifetimes are corrected by subtracting the bias due to the neglected $t$ \vs \sigmat correlations. The average absolute difference between the resulting value and the simulated lifetime, $0.24\fs$, is assigned as a systematic uncertainty due to the neglected background contamination in the \DzToKpi fit.

The background contamination under the \mbox{\DpToKpipi} peak is already accounted for in the fit of the \Dp lifetime using sideband data. In simulation, the sideband $(t,\sigmat)$ distribution describes the background $(t,\sigmat)$ distribution in the signal region well. The same might not hold in data given that some disagreement is observed between data and simulation in the $t$ distribution of the candidates populating the sideband. We fit to one thousand samples of simulated data obtained by sampling the fit PDF for the signal region and by resampling from the simulated \epem collisions for the sideband. The resulting samples feature sideband data that differ from the background in the signal region with the same level of disagreement as observed between data and simulation. The absolute average difference between the measured and simulated lifetimes, $2.52\fs$, is assigned as a systematic uncertainty due to the modeling of the background $(t,\sigmat)$ distribution. In the lifetime fit, the fraction of background candidates in the signal region is constrained from the fit to the \mDp distribution. When we change this background fraction to values obtained from fitting to the \mDp distribution with alternative signal and background PDFs, the change in the measured lifetime is negligible.

During data-taking, a periodic calibration determines the alignments and surface deformations of the internal components of the PXD and SVD and the relative alignments of the PXD, SVD, and CDC using \epem-collision, beam-background, and cosmic-ray events~\cite{Bilka:2020kgr}. Unaccounted-for misalignment can bias the measurement of the charmed decay lengths and hence their decay times. Two sources of uncertainties associated with the alignment procedure are considered: the statistical precision and a possible systematic bias. Their effects are evaluated using simulated signal-only decays reconstructed with a misaligned detector. For the statistical contribution, we consider configurations derived from comparison of alignment parameters determined from data acquired on two consecutive days. These configurations have magnitudes of misalignment comparable to the alignment precision as observed in data averaged over a typical alignment period. For the systematic contribution, we consider configurations derived from simulation studies in which coherent global deformations of the vertex detectors (\eg, radial expansion) are introduced~\cite{Brown_2009}. These deformations have magnitudes, determined by the most misaligned sensors, ranging from about 50\mum to 700\mum. The alignment procedure determines the magnitude of these deformations within 4\mum accuracy. We consider configurations in which the CDC is perfectly aligned and configurations in which it is misaligned. Possible effects on the determination of the IR are also introduced by using parameters measured on misaligned samples of simulated $\epem\to\mu^+\mu^-$ events, to fully mimic the procedure used for real data. For each misalignment configuration, we fit to the reconstructed signal candidates and estimate the lifetime bias. We estimate the systematic uncertainty due to imperfect detector alignment as the sum in quadrature of the largest biases observed in each of the statistical and systematic contributions. The resulting uncertainties are $0.72\fs$ and $1.70\fs$ for \DzToKpi and \DpToKpipi decays, respectively. The absolute length scale of the vertex detector is determined with a precision significantly better than $0.01\%$ and contributes negligibly to the systematic uncertainty.

The measurement of momenta is calibrated with the peak positions of abundant charmed-, strange-, and bottom-hadron decays. Uncertainty in the scaling of the momenta results in a systematic uncertainty in the lifetimes of 0.19\fs for \Dz and 0.48\fs for \Dp. Uncertainties in the \Dz and \Dp masses~\cite{pdg} contribute negligibly to the systematic uncertainty.

As a cross-check, a statistically independent measurement of the \Dz lifetime is performed using approximately $146\times10^3$ $\Dstarp\to\Dz(\to\Km\pip\pim\pip)\pip$ decays reconstructed in data with criteria similar to those used for the \DzToKpi mode and a signal purity greater than 99\%. The resulting lifetime, $\tauDzKpipipi\pm\tauDzStatKpipipi\stat\fs$, agrees with the value determined from the \DzToKpi mode.

Finally, the internal consistency of the measurement is tested by repeating the full analysis on various subsets of the data, \ie, running periods and running conditions, charmed-meson momentum and flight direction, and \Dstarp or \Dstarm candidates. We have also studied different selection criteria and sideband definitions. In all cases, the resulting changes in the lifetimes are insignificant.

In conclusion, the \Dz and \Dp lifetimes are measured using $\epem\to\c\cbar$ data collected by the \belletwo experiment corresponding to an integrated luminosity of \lumi. The results,
\begin{align}
\tau(\Dz) &= \phantom{0}\tauDz\pm\tauDzStat\stat\pm\tauDzSyst\syst\fs\quad\text{and}\\
\tau(\Dp) &= \tauDp\pm\tauDpStat\stat\pm\tauDpSyst\syst\fs\,,
\end{align}
are the world's most precise to date and are consistent with previous measurements~\cite{pdg}. Assuming that all systematic uncertainties are fully correlated between the two measurements, except those due to the background contamination (assumed uncorrelated), the total correlation coefficient is 18\%. The ratio of lifetimes is $\tau(\Dp)/\tau(\Dz) = 2.510\pm0.013\stat\pm0.007\syst$. These results demonstrate the vertexing capabilities of the \belletwo detector and confirm our understanding of systematic effects that impact future decay-time-dependent analyses of neutral-meson mixing and mixing-induced \CP violation. 
 
\begin{acknowledgments}
%
We thank the SuperKEKB group for the excellent operation of the
accelerator; the KEK cryogenics group for the efficient
operation of the solenoid; the KEK computer group for
on-site computing support; and the raw-data centers at
BNL, DESY, GridKa, IN2P3, and INFN for off-site computing support.
This work was supported by the following funding sources:
Science Committee of the Republic of Armenia Grant No.~20TTCG-1C010;
Australian Research Council and research Grants
No.~DP180102629,
No.~DP170102389,
No.~DP170102204,
No.~DP150103061,
No.~FT130100303,
No.~FT130100018,
and
No.~FT120100745;
Austrian Federal Ministry of Education, Science and Research,
Austrian Science Fund No.~P 31361-N36, and
Horizon 2020 ERC Starting Grant No.~947006 ``InterLeptons'';
Natural Sciences and Engineering Research Council of Canada, Compute Canada and CANARIE;
Chinese Academy of Sciences and research Grant No.~QYZDJ-SSW-SLH011,
National Natural Science Foundation of China and research Grants
No.~11521505,
No.~11575017,
No.~11675166,
No.~11761141009,
No.~11705209,
and
No.~11975076,
LiaoNing Revitalization Talents Program under Contract No.~XLYC1807135,
Shanghai Municipal Science and Technology Committee under Contract No.~19ZR1403000,
Shanghai Pujiang Program under Grant No.~18PJ1401000,
and the CAS Center for Excellence in Particle Physics (CCEPP);
the Ministry of Education, Youth, and Sports of the Czech Republic under Contract No.~LTT17020 and
Charles University Grants No.~SVV 260448 and No.~GAUK 404316;
European Research Council, Seventh Framework PIEF-GA-2013-622527,
Horizon 2020 ERC-Advanced Grants No.~267104 and No.~884719,
Horizon 2020 ERC-Consolidator Grant No.~819127,
Horizon 2020 Marie Sklodowska-Curie Grant Agreement No.~700525 "NIOBE",
and
Horizon 2020 Marie Sklodowska-Curie RISE project JENNIFER2 Grant Agreement No.~822070 (European grants);
L'Institut National de Physique Nucl\'{e}aire et de Physique des Particules (IN2P3) du CNRS (France);
BMBF, DFG, HGF, MPG, and AvH Foundation (Germany);
Department of Atomic Energy under Project Identification No.~RTI 4002 and Department of Science and Technology (India);
Israel Science Foundation Grant No.~2476/17,
U.S.-Israel Binational Science Foundation Grant No.~2016113, and
Israel Ministry of Science Grant No.~3-16543;
Istituto Nazionale di Fisica Nucleare and the research grants BELLE2;
Japan Society for the Promotion of Science, Grant-in-Aid for Scientific Research Grants
No.~16H03968,
No.~16H03993,
No.~16H06492,
No.~16K05323,
No.~17H01133,
No.~17H05405,
No.~18K03621,
No.~18H03710,
No.~18H05226,
No.~19H00682, %
No.~26220706,
and
No.~26400255,
the National Institute of Informatics, and Science Information NETwork 5 (SINET5), 
and
the Ministry of Education, Culture, Sports, Science, and Technology (MEXT) of Japan;  
National Research Foundation (NRF) of Korea Grants
No.~2016R1\-D1A1B\-01010135,
No.~2016R1\-D1A1B\-02012900,
No.~2018R1\-A2B\-3003643,
No.~2018R1\-A6A1A\-06024970,
No.~2018R1\-D1A1B\-07047294,
No.~2019K1\-A3A7A\-09033840,
and
No.~2019R1\-I1A3A\-01058933,
Radiation Science Research Institute,
Foreign Large-size Research Facility Application Supporting project,
the Global Science Experimental Data Hub Center of the Korea Institute of Science and Technology Information
and
KREONET/GLORIAD;
Universiti Malaya RU grant, Akademi Sains Malaysia, and Ministry of Education Malaysia;
Frontiers of Science Program Contracts
No.~FOINS-296,
No.~CB-221329,
No.~CB-236394,
No.~CB-254409,
and
No.~CB-180023, and No.~SEP-CINVESTAV research Grant No.~237 (Mexico);
the Polish Ministry of Science and Higher Education and the National Science Center;
the Ministry of Science and Higher Education of the Russian Federation,
Agreement No.~14.W03.31.0026, and
the HSE University Basic Research Program, Moscow;
University of Tabuk research Grants
No.~S-0256-1438 and No.~S-0280-1439 (Saudi Arabia);
Slovenian Research Agency and research Grants
No.~J1-9124
and
No.~P1-0135;
Agencia Estatal de Investigacion, Spain Grants
No.~FPA2014-55613-P
and
No.~FPA2017-84445-P,
and
No.~CIDEGENT/2018/020 of Generalitat Valenciana;
Ministry of Science and Technology and research Grants
No.~MOST106-2112-M-002-005-MY3
and
No.~MOST107-2119-M-002-035-MY3,
and the Ministry of Education (Taiwan);
Thailand Center of Excellence in Physics;
TUBITAK ULAKBIM (Turkey);
National Research Foundation of Ukraine, project No.~2020.02/0257,
and
Ministry of Education and Science of Ukraine;
the U.S. National Science Foundation and research Grants
No.~PHY-1807007 %
and
No.~PHY-1913789, %
and the U.S. Department of Energy and research Awards
No.~DE-AC06-76RLO1830, %
No.~DE-SC0007983, %
No.~DE-SC0009824, %
No.~DE-SC0009973, %
No.~DE-SC0010073, %
No.~DE-SC0010118, %
No.~DE-SC0010504, %
No.~DE-SC0011784, %
No.~DE-SC0012704, %
No.~DE-SC0021274; %
and
the Vietnam Academy of Science and Technology (VAST) under Grant No.~DL0000.05/21-23.
 \end{acknowledgments}

\ifthenelse{\boolean{wordcount}}%
{ \bibliographystyle{unsrt}
  \nobibliography{references}}
{ \bibliographystyle{apsrev4-1}
}

\end{document}